\documentclass[a4paper,fleqn,usenatbib]{mnras}

\usepackage[T1]{fontenc}
\usepackage{ae,aecompl}

\usepackage{graphicx}	
\usepackage{amsmath}	
\usepackage{amssymb}	
\usepackage{hyperref}
\usepackage{natbib}
\usepackage{upgreek}

\newcommand{\ergcms}{erg$\,$cm$^{-2}$$\,$s$^{-1}$\,}
\newcommand{\ergs}{erg$\,$s$^{-1}$\,}

\title[Neutron Star Hypothesis for Fomalhaut b]{A Test of the Neutron Star Hypothesis for Fomalhaut b}

\author[K. Poppenhaeger et al.]{
K. Poppenhaeger,$^{1, 2}$\thanks{E-mail: k.poppenhaeger@qub.ac.uk}
K. Auchettl,$^{3, 4}$
S.J. Wolk$^{2}$
\\
$^{1}$Astrophysics Research Centre, Queen's University Belfast, University Road, BT7 1NN Belfast, United Kingdom\\
$^{2}$Harvard-Smithsonian Center for Astrophysics, 60 Garden Street, Cambridge, 02138 MA, USA\\
$^{3}$Center for Cosmology and Astro-Particle Physics, The Ohio State University, 191 West Woodruff Avenue, Columbus, OH 43210, USA\\
$^{4}$Department of Physics, The Ohio State University, 191 West Woodruff Avenue, Columbus, OH 43210, USA
}

\date{Accepted 2017 March 1. Received 2017 February 28; in original form 2016 December 19.}

\pubyear{2017}

\begin{document}
\label{firstpage}
\pagerange{\pageref{firstpage}--\pageref{lastpage}}
\maketitle

\begin{abstract}

Fomalhaut b is a directly imaged object in the debris disk of the star Fomalhaut. It has been hypothesized to be a planet, however there are issues with the observed colours of the object that do not fit planetary models. An alternative hypothesis is that the object is a neutron star in the near fore- or background of Fomalhaut's disk. We test if Fomalhaut b could be a neutron star using X-ray observations with \textit{Chandra's} HRC-I instrument in the energy range of 0.08--10 keV. We do not detect X-ray emission from either Fomalhaut b or the star Fomalhaut itself. Our nondetection corresponds to an upper limit on the X-ray flux of Fomalhaut~b of $F_\mathrm{X} < 1.3\times10^{-14}$~\ergcms in the energy range 0.08--10~keV. For the A-type central star Fomalhaut, we derive an X-ray upper limit of $L_\mathrm{X} < 2.0\times10^{25}$~\ergs in the energy range 0.08--10~keV. Fomalhaut~b's X-ray non-detection constrains the parameter space for a possible neutron star significantly, implying surface temperatures lower than 91\,000~K and distances closer than 13.3~pc to the solar system. In addition we find that reflected starlight from the central star fits the available optical detections of Fomalhaut~b; a smaller planet with a large ring system might explain such a scenario.

\end{abstract}



\section{Introduction}

The very first successfully imaged exoplanet is Fomalhaut b \citep{Kalas2008}. The Fomalhaut system consists of a young main-sequence star of spectral type A4, Fomalhaut A, and there are two other physically bound stars in this system, Fomalhaut B, which is of spectral type K, and Fomalhaut C, which is of spectral type M. Both are located at very wide distances from the primary star, at ca.\ 100'' corresponding to ca.\ 750~AU. Fomalhaut A is located at a distance of 7.5~pc from the Sun. It is surrounded by a debris disk that has been imaged in the optical and in the infrared. Repeated observations performed with the coronograph onboard \textit{HST} at red optical wavelengths revealed an object that appeared to be moving on an orbit in the same plane as the debris disk with an orbital distance of ca.\ 100~AU; this object was called Fomalhaut b. An exoplanet at that distance to the host star is expected to be cold, and respective models yielded that the object likely has a mass roughly similar to that of Jupiter \citep{Kalas2008}. 

However, observations at other wavelengths yielded surprising results. The planet was repeatedly detected at optical and near-infrared wavelengths of around 6000 and 8100\,\AA~\citep{Kalas2008, Currie2012}, but observations at even longer wavelengths, where the exoplanet is expected to be brighter, yielded strong upper limits that were incompatible with typical exoplanetary model spectra \citep{Janson2012}. Even more confusingly, Fomalhaut b was detected at blue optical wavelengths (ca.\ 4400\,\AA), where it should have been very dim \citep{Currie2012}. Recent detections at optical wavelengths also revealed that the best-fit orbit of this object is not, as previously thought, aligned with the debris disk; instead, the object seems to be on a disk-crossing orbit \citep{Kalas2013}. Several possible explanations have been brought forward to explain the surprising spectral shape of this object, including a planetary accretion disk around a Jupiter-mass planet, explaining the blue excess by reflection, but not explaining the out-of-disk orbit \citep{Kalas2008, Janson2012}; a small hot planet with a very large cloud of planetesimals, explaining the brightness and misaligned orbit without producing unobserved disk perturbations, but not explaining how it arrived there since no scattering objects of sufficient mass are available \citep{Janson2012}; and an icy dust cloud without a core, explaining the brightness, but being implausible in terms of age estimates \citep{Currie2012, Galicher2013}. None of these hypotheses is fully consistent with the object properties observed so far.

\begin{figure}
\includegraphics[width=0.48\textwidth]{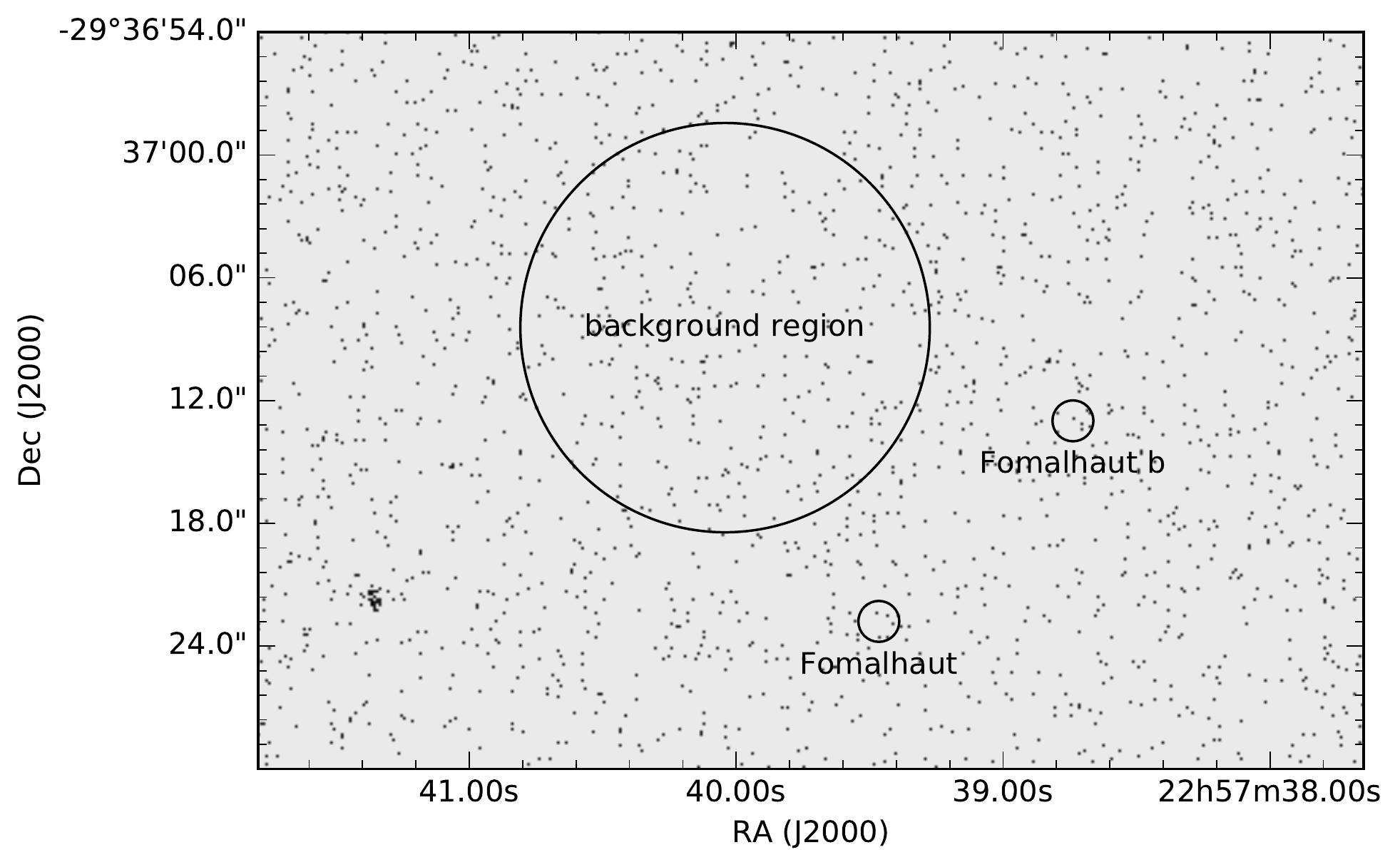}
\caption{The X-ray image collected with HRC-I shows that the positions of Fomalhaut and Fomalhaut~b (marked with small circles) do not display any significant photon excess over the background (large circle).}
\label{hrcimagezoomed}
\end{figure}

Given that the latest observations of relative motion observed between Fomalhaut A and Fomalhaut b do not support an in-disk orbit of the planet candidate, one should consider the possibility that Fomalhaut b might be a background object that just happens to pass by Fomalhaut~A at a close projected distance. This has recently been discussed by \cite{Neuhaeuser2015}.

The faintness of Fomalhaut b at optical wavelengths (ca.\ 25 mag) and the blue-leaning SED constrain what kind of object it can be. It cannot be an ordinary star, since any star with such a blue SED would be intrinsically bright. To match Fomalhaut b's faintness, the hypothetical star would have to be so far away that substantial reddening caused by the interstellar medium would occur, so that the colors would not fit any more. It also cannot be a white dwarf; white dwarfs typically have absolute visible magnitudes of 10--15 mag \citep{WoodOswalt1998}. Placing a white dwarf at a distance of 0.5--5 kpc to match the brightness would mean that the projected 2D spatial velocity of the white dwarf would have to be unrealistically large in order to fit the observed trajectory of Fomalhaut~b \citep{Neuhaeuser2015}.

This leaves the possibility that Fomalhaut b may be a neutron star. Neutron stars have absolute visual magnitudes of ca.\ 22--25 mag, depending on temperature \citep[for example]{Kaplan2011}. The apparent magnitude of Fomalhaut b is ca.\ 25 in the visual, so an old and cool neutron star at ca.\ 10 pc distance or a somewhat hotter one farther away would generally fit the observed apparent brightness of the object. Neutron stars start out with surface temperatures of several MK after formation and cool over time scales of megayears to temperatures of around 100\,000 K  \citep{Page1992, Yakovlev2004review}. Several nearby thermally emitting neutron stars have been detected in X-rays with sufficient signal to noise to allow spectral fitting, and those neutron stars typically display surface temperatures of 500\,000 to 1\,000\,000~K; however, they are estimated to be rather young with $\lesssim 1$~Myr \citep{Kaplan2011}. An older, cooler neutron stars will still look blue at optical wavelengths, matching the few observational detections available so far for Fomalhaut~b. The emission from an old neutron star would peak in the far-UV, however the brightness of the central star Fomalhaut A at optical and UV wavelengths makes such observations difficult and requires sophisticated subtraction methods for the stellar point spread function (PSF). However, in soft X-rays, where the hard tail of the neutron star spectrum may be detectable, a star with the mass of Fomalhaut is expected to be X-ray dark, making X-ray observations a good opportunity to study the object Fomalhaut~b.

\section{Observations and data analysis}

The Fomalhaut system has been imaged in X-rays previously, albeit spatially unresolved, with the \textit{Einstein} and \textit{ROSAT} observatories. The deepest archival observation was performed with a \textit{ROSAT} PSPC pointing in 1992, and we present that data set shortly in the following section. We have performed a new, deeper X-ray observation which can spatially resolve Fomalhaut and Fomalhaut~b using \textit{Chandra}'s HRC-I instrument and present that data in detail in section~\ref{newobs}.

\subsection{The archival \textit{ROSAT} pointing}

\textit{ROSAT} was a space telescope that operated from 1990 to 1999, observing the soft X-ray sky \citep{Truemper1982, Aschenbach1988}. Its PSPC camera \citep{Pfeffermann1987} provided soft X-ray images with a spatial resolution of 25 arsec (full-width half maximum of the point spread function) in an energy range of 0.1--2.4~keV. The Fomalhaut system was observed in a PSPC pointing for 6.3~ks in 1992. 
The spatial resolution of \textit{ROSAT}'s PSPC is not sufficient to resolve the Fomalhaut system since the spatial separation between Fomalhaut and Fomalhaut~b is 13.8 arcsec.

The Fomalhaut system is undetected in X-rays by \textit{ROSAT}. 
\cite{Schmitt1997} give an upper limit on the X-ray count rate of 
0.0066~cts/s for this observation. In the interest of a consistent 
analysis of the \textit{ROSAT} and \textit{Chandra} data, we 
re-analyzed the data set and derive a 3$\sigma$ upper limit on 
the count rate of 0.008~cts/s. The flux that corresponds to this 
upper limit depends on the assumed spectrum of the undetected source, 
and we discuss this in more detail in section~\ref{results}. Assuming 
a black-body spectrum with a temperature of 100,000~K, we calculate 
an upper limit on the X-ray flux of $9.0\times 10^{-14}$~erg\,cm$^{-2}$\,s$^{-1}$. 
We used the WebPIMMS 
count rate simulator for this \citep{Mmukai1993}, which is accessible 
at \url{http://heasarc.gsfc.nasa.gov/cgi-bin/Tools/w3pimms/w3pimms.pl}. We took into
account that the Boron filter was used during this \textit{ROSAT} observations, 
which reduces the throughput at soft X-ray energies by ca.\ 90\%.

\subsection{The new \textit{Chandra} observation}\label{newobs}

We performed an X-ray pointing with \textit{Chandra}'s HRC-I \citep{Weisskopf2000} with a duration of 32.7~ks. Details of the observation are given in Table~\ref{xrayobs}. We used the standard data analysis tools of \textit{Chandra}'s \textit{CIAO} software, version 4.7 \citep{ciao}. Specifically, we extracted counts from a source detect cell placed on the nominal positions of Fomalhaut and Fomalhaut~b, each with a radius of $1^{\prime\prime}$. This was compared to a background signal which was extracted from a source-free region on the chip with a radius of $20^{\prime\prime}$. Since HRC-I provides no useful energy resolution for the collected X-ray photons (in contrast to other X-ray instruments onboard \textit{Chandra}), no energy filtering was performed. The extracted photons therefore can come from the full range of energies to which HRC-I is sensitive, i.e.\ 0.08--10~keV.

We show the extracted image collected with HRC-I in 
Fig.~\ref{hrcimagezoomed}. 
The indicated regions show the positions of Fomalhaut and Fomalhaut~b, 
as well as our chosen background region. The background region contains 
294 photon counts, while the source detect cells for Fomalhaut and Fomalhaut~b contain
6 and 7 photon counts, respectively. Given that the background extraction region has an area
100 larger than the individual source detect cells, we would expect a pure background signal of
2.94 in a source detect cell.

\begin{table}
\caption{X-ray observation parameters for Fomalhaut~b.}
\label{xrayobs}
\begin{tabular}{l l}
\hline \hline
Instrument     & \textit{Chandra} HRC-I\\ \hline
Observation ID & 17896 \\
Exposure time  & 32691.2 s\\
Start time     & 2016-08-09 10:38:42 \\
End time       & 2016-08-09 20:22:24\\
Energy range   & 0.08--10 keV\\
\hline
\end{tabular}
\end{table}

\begin{table*}
\caption{Observed fluxes and upper limits for Fomalhaut~b in different wavelength bands.}
\label{allobs}
\begin{tabular}{l l l l l l l}
\hline \hline
Obs.\ year & Telescope/Instr. & Filter & Detection & flux & flux density  & Reference \\
& & & & [mag] & [erg\,cm$^{-2}$\,s$^{-1}$\,\AA$^{-1}$] & \\
\hline
1992      & ROSAT PSPC        & Boron filter & no  & ...            & $< 7.55\times10^{-16}$ & this work \\
2004      & HST ACS/HRC       & F606W        & yes & $24.92\pm0.10$ & $3.14\pm0.29\times10^{-19}$ & Currie et al.\ (2012) \\
2004      & HST ACS/HRC       & F606W        & yes & $24.42\pm0.09$ & $4.93\pm0.41\times10^{-19}$ & Kalas et al.\ (2008) \\
2004      & HST ACS/HRC       & F606W        & yes & $24.29\pm0.08$ & $5.61\pm0.41\times10^{-19}$ & Kalas et al.\ (2008) \\
2005      & KeckII NIRC2      & H            & no  & $\geq22.9$ & $\leq 8.28\times10^{-20}$ & Kalas et al.\ (2008) \\
2005      & KeckII NIRC2      & CH$_4$S      & no  & $\geq20.6$ & ... & Kalas et al.\ (2008) \\
2006      & HST ACS/HRC       & F435W        & no  & $\geq24.7$ & $8.36\times10^{-19}$ & Kalas et al.\ (2008) \\
2006      & HST ACS/HRC       & F435W        & yes & $25.22\pm0.18$ & $5.18\pm0.86\times10^{-19}$ & Currie et al.\ (2012) \\
2006      & HST ACS/HRC       & F606W        & yes & $25.13\pm0.09$ & $2.59\pm0.21\times10^{-19}$ & Kalas et al.\ (2008) \\
2006      & HST ACS/HRC       & F606W        & yes & $24.97\pm0.09$ & $3.00\pm0.25\times10^{-19}$ & Currie et al.\ (2012) \\
2006      & HST ACS/HRC       & F814W        & yes & $24.55\pm0.13$ & $1.69\pm0.20\times10^{-19}$ & Kalas et al.\ (2008) \\
2006      & HST ACS/HRC       & F814W        & yes & $24.91\pm0.20$ & $1.21\pm0.22\times10^{-19}$ & Currie et al.\ (2012) \\
2008      & Gemini North NIRI & L$^\prime$   & no  & $\geq 16.6$ & $\leq 1.22\times10^{-18}$ & Kalas et al.\ (2008) \\
2009      & Subaru IRCS       & J            & no  & $\geq 22.22$ & $\leq 4.01\times10^{-19}$ & Currie et al.\ (2012) \\
2010/2011 & Spitzer IRAC      & 4.5\,$\upmu$m  & no  & $\geq 16.7$ & $\leq 5.71\times10^{-19}$ & Janson et al.\ (2012) \\
2013      & Spitzer IRAC      & 4.5\,$\upmu$m  & no  & $\geq 17.3$ & $\leq 3.30\times10^{-19} $ & Janson et al.\ (2015) \\
2016      & Chandra HRC-I     & 0.08-10 keV  & no  & ... & $< 8.58\times10^{-17}$ & this work \\
\hline
\end{tabular}
\end{table*}

Using Poissonian counting statistics, a signal of
9~counts 
in a detect cell would correspond to a $\geq 3\sigma$ excess, 
i.e.\ a probability smaller than 0.03\% that this signal would 
be detected as a random fluctuation of the background. However, 
the signal in the detect cells is lower than that, meaning that 
a significant count excess is not detected for either 
Fomalhaut or Fomalhaut~b.
\footnote{In contrast, the source to the lower left of the background region in 
Fig.~\ref{hrcimagezoomed} has 33 photon counts in a $1^{\prime\prime}$
detect cell and is significantly detected. No matches were found 
for this source position  in major catalogs at various wavelengths, 
so that we can make no further 
statements about the nature of this source.}
We therefore adopt a signal of 
9~counts
counts and, given the exposure time of 32691.2~s, a count rate of
0.00028~cts/s 
as the upper limit for Fomalhaut and Fomalhaut~b in the full HRC energy
band of 0.08--10~keV.

What we are specifically interested in is an upper limit on the X-ray flux and X-ray luminosity of Fomalhaut~b and Fomalhaut. Often this is done by spectral fitting to model the energy sensitivity of the detector and derive the actual source flux. However, since the \textit{{HRC-I}} instrument has virtually no intrinsic energy resolution, one needs to find an appropriate counts-to-energy conversion factor for the observation. This conversion factor will depend on the underlying X-ray spectrum one assumes for a source, because \textit{{HRC-I}} has different sensitivities to photons at the upper and lower end of the energy range it is sensitive to. To give the reader a rough idea, the effective area of \textit{{HRC-I}} starts at 0.08~keV with a few cm$^2$ and rises in several steps to ca.\ 200~cm$^2$ at 1~keV (with a narrow downward spike around 0.3~keV due to an absorption edge of carbon residuals on the mirror assembly), and decreases again at energies higher than 1~keV. Therefore a spectrum peaking at soft energies where the effective area is small will lead to a small number of counts being detected, whereas a spectrum with the same flux, but peaking at harder X-ray energies where \textit{{HRC-I}} has a larger effective area, will lead to more counts being detected.

\section{Results}\label{results}

\subsection{Fomalhaut b as a neutron star}

Neutron stars emit strongly in the far-UV and in soft X-rays. Following \citet{Neuhaeuser2015}, we will now explore the hypothesis of Fomalhaut~b being a neutron star in the near fore- or background of the Fomalhaut disk, given the new observations we performed with \textit{Chandra}.

\subsubsection{The X-ray flux upper limit}

To determine the appropriate upper limit to the X-ray flux of Fomalhaut~b from the upper limit on the X-ray count rate, one needs to assume some sort of underlying hypothetical source spectrum. Neutron stars are generally soft X-ray sources \citep{Truemper2005, Haberl2007}. Soon after their formation their surface cools down to a few MK, and reaches temperatures of a few $100,000$~K and lower after several million years \citep{Yakovlev2004review}. 

We can derive an appropriate count-to-energy conversion factor for \textit{Chandra} HRC-I by using a blackbody spectral model in WebPIMMS. Using a hotter neutron star model with a surface temperature of 1~MK over an X-ray energy range of 0.08--10~keV yields a conversion factor of $8.18\times 10^{-12}$~ergs\,cm$^{-2}$\,s$^{-1}$\,count$^{-1}$ for HRC-I; using a colder neutron star model with $100\,000$~K yields a conversion factor of $4.79\times 10^{-11}$~ergs\,cm$^{-2}$\,s$^{-1}$\,count$^{-1}$, i.e.\ larger by a factor of about five. An upper limit on the X-ray flux will therefore depend on the assumed surface temperature of the neutron star. 

Fortunately, this dependence and the actual fitting of possible spectral models including the optical data are `self-correcting': Say we assume a surface temperature that is too high. This will lead to a small conversion factor, meaning a very low, i.e.\ very restrictive, upper limit on the X-ray flux. This in turn will force us to fit the data with a rather cool neutron star model, because a hot neutron star would be too bright in X-rays and therefore violate the upper limit. The mismatch between our initial assumption for the surface temperature and the derived fit will indicate that the assumption was not correct. Similarly, assuming a neutron star that is too cold will lead to a large conversion factor, a non-restrictive upper limit on the X-ray flux, and therefore allow the data to be fit with a hotter neutron star. Finally, a roughly correct assumption about the surface temperature will lead to a matching temperature in the fit.

Having gone through this exercise, we assume a cool 
surface temperature of $100\,000$~K for Fomalhaut~b as a 
neutron star and use a counts-to-energy conversion factor 
of $4.79\times 10^{-11}$~ergs\,cm$^{-2}$\,s$^{-1}$\,count$^{-1}$. 
The upper limit on the count rate of 
0.00028~cts/s 
therefore corresponds to an upper limit on the X-ray flux of
$1.3\times 10^{-14}$~ergs\,cm$^{-2}$\,s$^{-1}$. We calculate the upper limit on the average flux density for 
the wavelength range covered by \textit{HRC-I}, i.e.\ 0.08-10~keV, to be 
$8.6\times 10^{-17}$~ergs\,cm$^{-2}$\,s$^{-1}$\,\AA$^{-1}$.

\begin{figure*}
\includegraphics[width=0.48\textwidth]{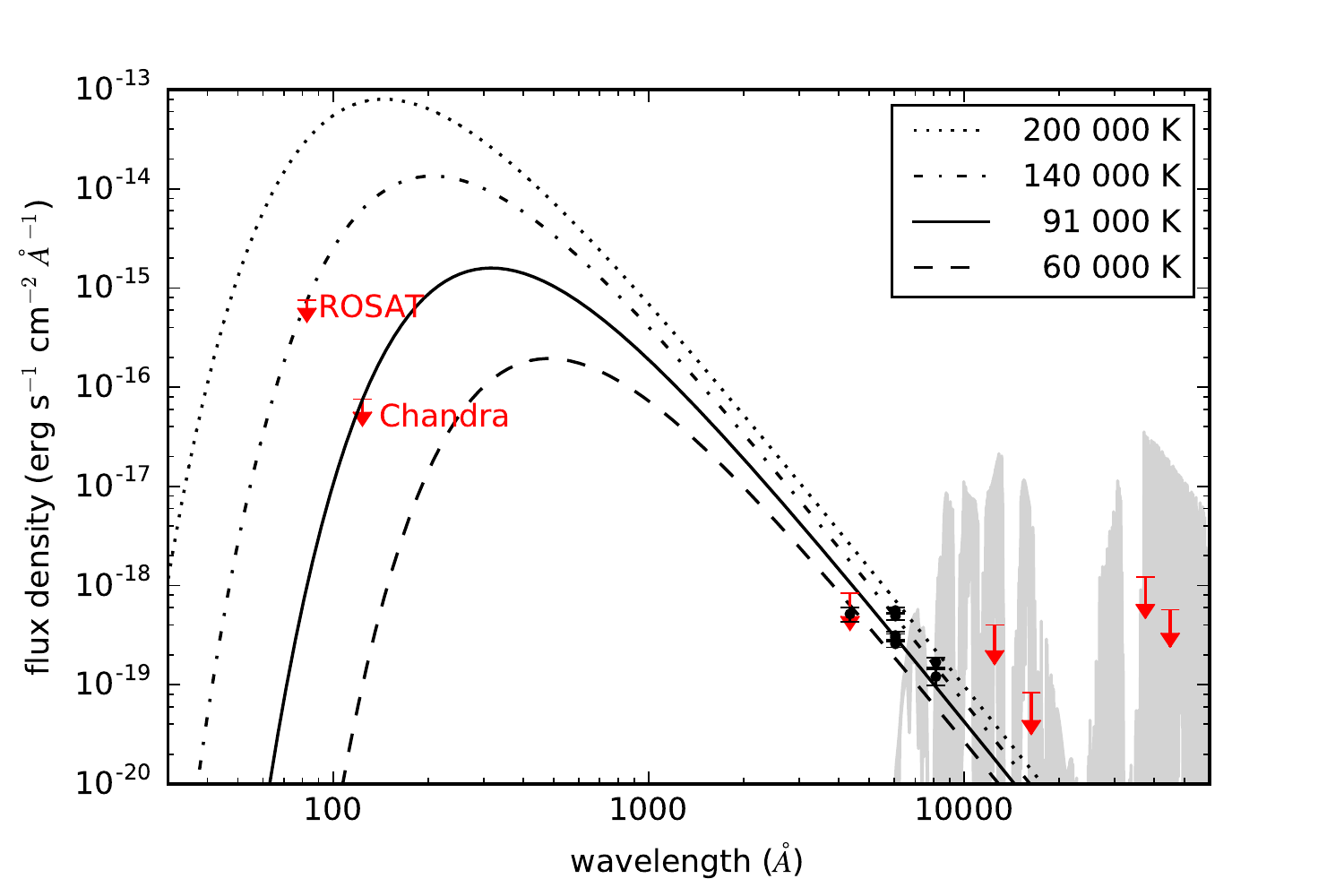}
\includegraphics[width=0.48\textwidth]{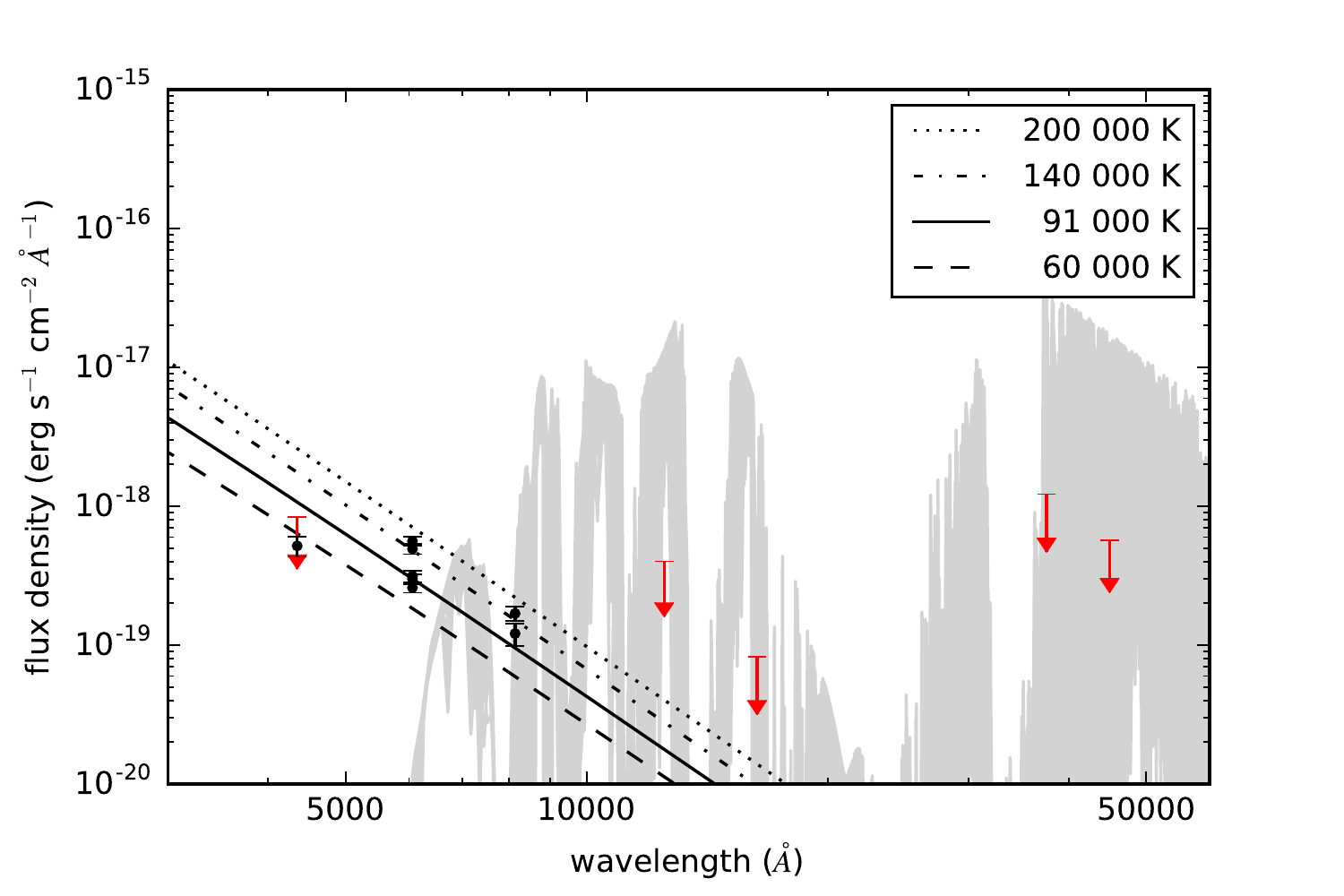}

\caption{The detections (black) and upper limits (red) for Fomalhaut b, together with an expected spectrum for a 400~K cold giant planet in light grey (from the pre-computed model grids available at \url{https://phoenix.ens-lyon.fr/Grids/}; these spectra are based on the PHOENIX code \citep{Hauschildt1997} with adaptations for sub-stellar objects \citep{Chabrier2000, Allard2001}) and blackbody spectra for a cool neutron star with temperatures of 200\,000~K, 140\,000~K, 91\,000~K, and 60\,000~K at a distance of 13.5~pc. The plot to the right shows a zoom-in onto the optical and infrared part of the spectrum. A neutron star with $T=91\,000$~K and a distance of 13.5~pc is the best-fitting neutron star model that fits the optical/near-IR detections and grazes the X-ray upper limit. }
\label{spectra}
\end{figure*}
\subsubsection{Optical and near-infrared data}

Fomalhaut~b has been detected by several groups in the optical and the near-infrared. Some detections in the optical have been reported to have significantly differing fluxes when taking the reported error bars at face value; however, given the sophisticated image analysis necessary to detect Fomalhaut~b's emission at all, we consider it is possible that the actual flux errors may be larger than initially reported. Upper limits at longer wavelengths have been derived from observations as well. \citet{Neuhaeuser2015} have given a summary on available pre-2015 detections and upper limits, and we give an updated overview of the available data in Table~\ref{allobs}, including a new data point in the infrared by \citet{Janson2015} and the X-ray flux upper limits derived in this work.

We can use those data sets to constrain an assumed neutron star spectrum for Fomalhaut~b. One might in principle assume a reddening effect on the spectrum occurs from dust absorption along the line of sight between Fomalhaut~b and the observer. However, interstellar absorption is negligible for nearby objects. Dust absorption from the Fomalhaut disk is also unlikely to produce significant reddening: \cite{Marsh2005} observed the Fomalhaut disk at 350~$\upmu$m and modelled a reddening of 0.005~mag at a wavelength of 24~$\upmu$m from those observations. From this we can extrapolate, using appropriate conversions from the 24~micron band to the $K$ and $V$ bands \citep{Indebetouw2005, Xue2016}, that the reddening in the $V$ band will be ca.\ 0.17~mag. This is smaller than the rms scatter of the Fomalhaut~b detections in the observed optical bands, which is of the order of 0.3~mag (see Table~\ref{allobs}), and we therefore regard a potential dust reddening due to the Fomalhaut disk as irrelevant to our further analysis.
\subsubsection{Possible neutron star parameters for Fomalhaut~b}

The emission from a neutron star in its simplest form can be described by a blackbody spectrum. It has been shown that often neutron stars are more accurately described by adding an additional harder X-ray component to the spectrum. Physically, this is likely due to hot polar caps covering a certain fraction of the neutron star surface (see \citet{Becker1997} and references therein) or the presence of a pulsar wind nebula \citep{Stappers2003}; Those harder X-ray spectral components are typically described by a power-law or a thermal spectrum. However, as the simplest possible model, we are assuming a single temperature blackbody component for the neutron star surface. Furthermore, we assume a typical radius of the hypothetical neutron star of 10~km \citep{Oezel2016review}. Any discrepancy to this assumed radius will only cause a slightly smaller or larger distance to the object in the spectral fit. 

We produce a least-squares fit of a blackbody model to the available data, using the detections of Fomalhaut~b at optical wavelengths and treating the Chandra upper limit as a detection data point for the sake of the fit. The other upper limits in the optical and IR do not restrict the fit (see Figure~\ref{spectra}).

The best fit is found for a blackbody temperature of 91\,000~K and a distance of 13.5~pc. The constraint on the temperature is driven by the X-ray upper limit, while the detections in the optical and near-infrared pinpoint the distance for a given temperature. We indicate the effect of changing the temperature for a given distance (the latter being interchangeable with the radius of the object) in Figure~\ref{spectra}; a closer distance moves the spectrum upwards, while a cooler temperature moves the spectrum to the lower right according to Wien's displacement law. This means that one cannot assume a neutron star hotter than 91\,000~K and place it at a larger distance, because such a model would either lie below the optical data points while satisfying the X-ray upper limit, or be a reasonable fit to the optical data while violating the X-ray upper limit. Note that the best-fitting neutron star model is found for the case where the object is indeed detected at the X-ray upper limit, which it is not. Therefore, if Fomalhaut~b is a neutron star, it has to be cooler than 91\,000~K and closer than 13.5~pc. 

Additionally, using the largest distance estimate of 13.5~pc, we can convert the upper limit to Fomalhaut~b's X-ray flux into an upper limit to its X-ray luminosity. We calculate this upper limit to be $L_\mathrm{X} < 2.7\times10^{26}$~erg/s.

\subsection{The A-type star Fomalhaut}

The central star Fomalhaut is an intermediate-mass star of spectral type A4V. Massive stars (spectral types O and early B) drive winds which produce X-rays through shocks, while low-mass stars (mid-F to late M) emit in X-rays due to their magnetically heated coronae. Stars in the intermediate mass regime, i.e. of spectral types mid-B to early F, typically do not show X-ray emission, with the exception of the chemically peculiar Ap/Bp stars. Among regular, i.e.\ non-chemically-peculiar A-type stars, only a handful have been detected in X-rays. Those rare examples display weak and very soft X-ray emission that can be described by a coronal plasma with dominant temperature components around 1~MK in the case of the A5V star beta Pictoris \citep{Guenther2012betaPic} and temperature components in the range from 1-4~MK in the case of the A7V star Altair \citep{Robrade2009Altair}. Other X-ray observations of A-type stars have not led to detections \citep{Pease2006, Ayres2008Vega, Drake2014}, with the exception of HR~8799, which is listed by SIMBAD as an A6 star, but was shown by \cite{Robrade2010HR8799} to be a peculiar type of early-F star with coronal emission, therefore not being representative for the intermediate-mass stellar regime.

In our observation, Fomalhaut is undetected in X-rays, 
with an upper limit to the count rate of 0.00028~cts/s. Using a soft coronal 
plasma model with a temperature of 1~MK to derive the appropriate 
counts-to-energy conversion factor ($1.10\times10^{-11}$~erg/s/count), this upper 
limit corresponds to an upper limit to the X-ray flux of $F_\mathrm{X} < 3.03\times10^{-15}$~erg\,s$^{-1}$\,cm$^{-2}$ 
and, using Fomalhaut's distance of 7.7~pc, an X-ray luminosity of $L_\mathrm{X} < 2.0\times10^{25}$~erg/s
over the energy range of 0.08--10~keV. Converting this to the more regularly used 
energy range of 0.2--2~keV, we derive an upper limit of 
$L_\mathrm{X,\, 0.2-2} < 6.3\times10^{24}$~erg/s.

The ratio of X-ray and bolometric luminosity, often used 
as an activity indicator for stars that can display magnetic activity 
\citep{Telleschi2007, Poppenhaeger2010, Wright2011}, 
is $L_\mathrm{X}/L_\mathrm{bol} < 3.2\times10^{-10}$ for the energy range of 0.08--10~keV 
and $L_\mathrm{X,\, 0.2-2}/L_\mathrm{bol} < 9.9\times10^{-11}$ for the energy range of 
0.2--2~keV; a bolometric luminosity of $16.63$ times the solar bolometric luminosity, 
i.e.\ $6.36\times10^{34}$\,\ergcms \citep{Mamajek2012, Davis2005} was adopted for 
Fomalhaut. Fomalhaut is therefore a star for which magnetic activity is 
present at vanishingly low levels, if at all.

\begin{figure*}
\includegraphics[width=0.48\textwidth]{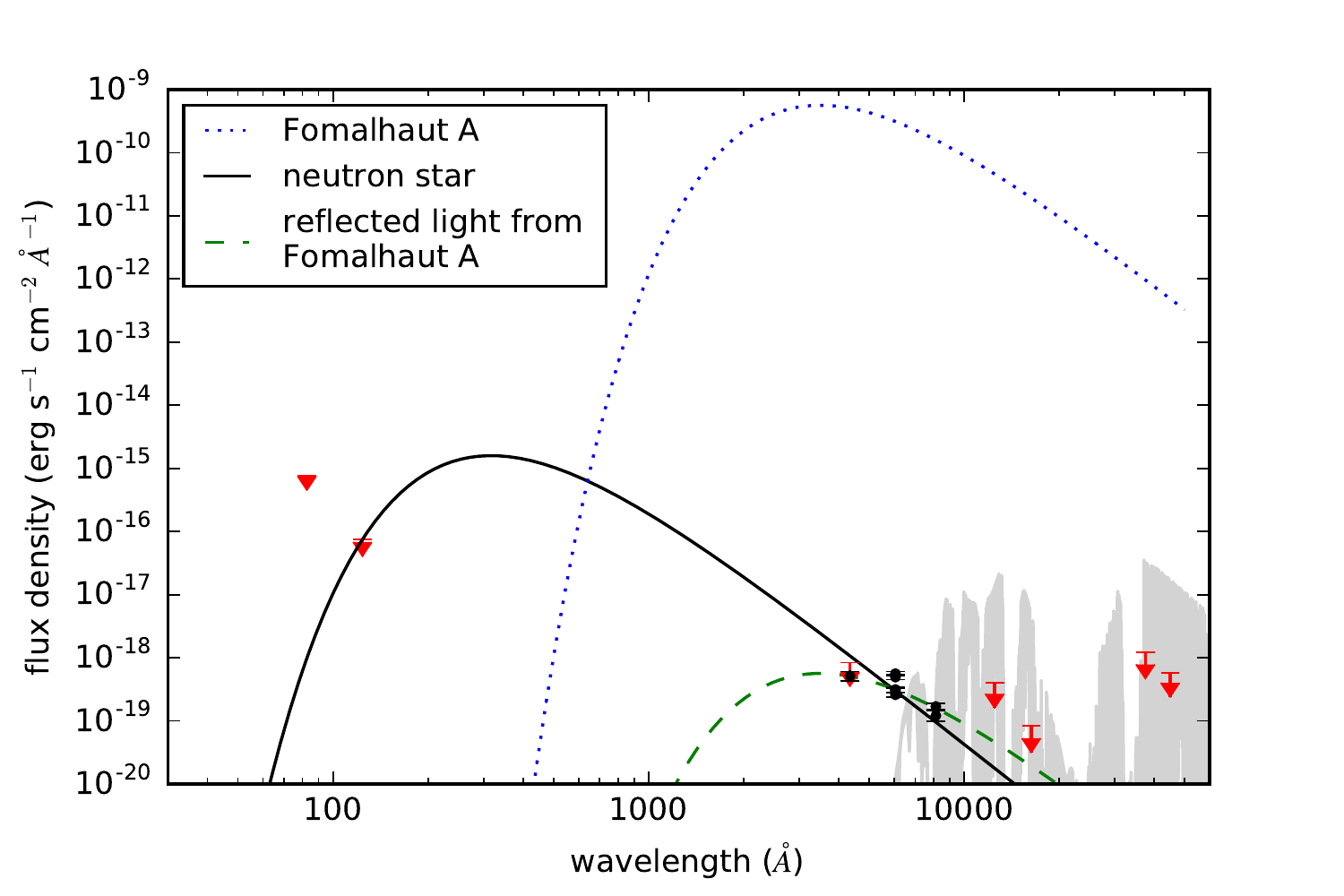}
\includegraphics[width=0.48\textwidth]{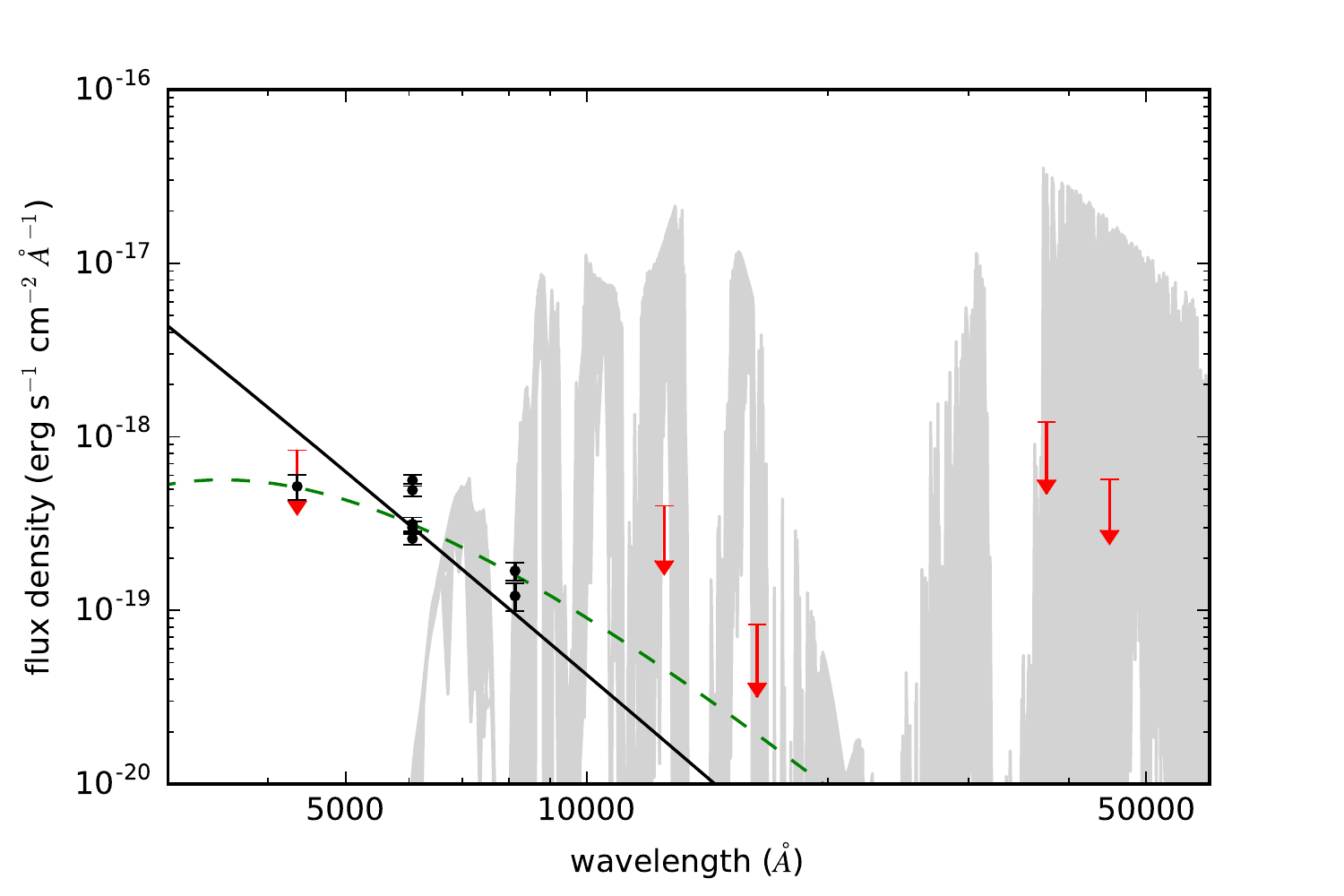}

\caption{Left: The emission from the central A4V star Fomalhaut~A dominates at wavelengths below 100~\AA (blue dotted line). Right: Both a cool neutron star and reflected starlight, i.e.\ a scaled down spectrum of Fomalhaut~A, could explain the optical detections for Fomalhaut~b (black solid line and green dashed line). }
\label{Astarspectra}
\end{figure*}
To set this into context, this is a very restrictive upper limit, similar to the upper limit derived with \textit{Chandra}'s ACIS and HRC instruments for the A0V star Vega, which has been observed to be $L_\mathrm{X} \leq 2\times 10^{25}$~erg/s \citep{Pease2006, Ayres2008Vega}. 
\section{Discussion}
Our non-detection of Fomalhaut~b in X-rays places strong constraints on the neutron star hypothesis. In order to fit the available data with a single-temperature blackbody model and a typical neutron star radius of 10~km, the hypothetical neutron star would need to be colder than ca.\ 90\,000~K, i.e.\ it would need to be very old with $\gtrsim 10$~Myr according to typical cooling models \citep{Yakovlev2004review}, and need to be located at a distance closer than 13.5~pc.\footnote{As already argued by \citet{Neuhaeuser2015}, any accompanying supernovae remnant from the formation of the neutron star would have diffused after ca.\ $100\,000$~years, consistent with an old age of the hypothetical neutron star.} This would make Fomalhaut~b the nearest known neutron star to the solar system, and furthermore an extremely old and cool one. If the neutron star radius was very small, for example 2~km, the temperature restriction would still hold, but the upper limit on its distance would only be ca.\ 0.5~pc. To set this into context, we can compare Fomalhaut~b to the extremely cool and old neutron star reported by \citet{Keane2013}. They derive an upper limit on temperature and black-body luminosity in the soft X-ray band of $<280\,000$~K and $<4.3\times10^{30}$~erg/s for the neutron star J1840--1419, which is located at a distance of 900~pc. Obviously, the close distance to the hypothetical neutron star for Fomalhaut~b is a main driver of the very low upper limit we have derived. However, it is still noteworthy that this would be the most near-by and coldest neutron star known, i.e.\ even with a neutron star explanation for Fomalhaut~b's surprising colours, we would be facing a very unusual scenario for \textit{detected} neutrino stars. It is worth pointing out that in general one can expect old and cool neutron stars to be quite common from supernova occurrence rates in the Milky Way (see for example \citealt{Camenzind2007book} p.269). However, since they are dim objects at all wavelengths, detecting those old and cool neutron stars is extremely challenging; if Fomalhaut~b truly is a neutron star, it would be the coldest one detected so far.

While a neutron star scenario is not completely ruled out, it is worthwhile to compare it to a scenario in which the detected emission of Fomalhaut~b is reflected light from the central star, and the reflecting object would be an object residing within the Fomalhaut disk (see for example \citet{Kalas2008}). As a rough guideline, we show a blackbody spectrum with appropriate parameters to approximate an A4V star like Fomalhaut ($T_\mathrm{eff} = 8270$~K \citep{Pecaut2013} and $L_\mathrm{bol} = 16.63 L_\mathrm{bol,\,\odot}$ \citep{Mamajek2012}, and from that calculated $R_\ast = 1.38\times 10^{11}$~cm) in Fig.~\ref{Astarspectra} (left-hand side). Scaling down this spectrum to account for a smaller reflecting surface at the position of Fomalhaut~b, we find a good match to the optical data when scaling down by a factor of $10^{-9}$, as shown in Fig.~\ref{Astarspectra} (right-hand side). Indeed, the reflected starlight hypothesis matches the optical data point in the blue ($435$~\AA) and in the near-infrared ($814$~\AA) better than the neutron star models which over-estimate the blue part and under-estimate the near-infrared part. We can perform a rough estimate of what size such a reflecting object would need to have to yield this observed flux. Assuming the reflector is in the Fomalhaut disk (i.e.\ at a distance of 115~AU to the central star) and has perfect reflectivity, we know that in order to reflect $10^{-9}$ of the starlight the reflector needs to have $10^{-9}$ times the area of a sphere with a radius of $115$~AU (as this sphere would encompass the complete flux of Fomalhaut). This corresponds to an area of ca.\ $4\times10^{12}$\,km$^2$. For comparison, the area enclosed by Saturn's outermost rings is ca.\ $6\times10^{10}$\,km$^2$. Much larger ring systems around planets may exist; the system 1SWASP J140747.93-394542.6 has been recently reportet to undergo extreme brightness changes which can be explained by the presence of a large ring system with a maximum radius of 0.6~AU \citet{Mamajek2012, Kenworthy2015}. That ring system would enclose an area of ca.\ $2.5\times10^{16}$\,km$^2$. For the Fomalhaut system, it is therefore conceivable that a smaller planet with a relatively large ring system with an area of $4\times10^{12}$\,km$^2$, i.e.\ a radius of ca.\ 600,000~km, could reflect enough starlight to match the observations.

To distinguish between these two hypotheses, imaging of Fomalhaut~b at far-UV or EUV wavelengths would be more definitive, however difficult since there are currently no telescopes in operation at those wavelengths. In the wavelength range of 200--400~\AA\ the neutron star and reflected light models begin to significantly differ from each other (compared to the optical); unfortunately, this is also the spectral range where the emission from the central star is strongest. In the EUV around 100~\AA\ a cool neutron star would display its strongest emission, while the central star is already relatively faint.

\section{Conclusion}

We have presented new X-ray observations of the Fomalhaut system that can spatially resolve the position of the central A4V star and the suspected planet Fomalhaut~b. We report non-detections of both the central star and the suspected planet in X-rays. 
The non-detection of the central star places a strict upper limit on its X-ray flux and its fractional X-ray luminosity $L_\mathrm{X}/L_\mathrm{bol} < 10^{-10}$ for the energy range of 0.2--2~keV, comparable to the best X-ray upper limits which have been obtained for A-type stars, albeit in a small sample. 

The hypothesis that Fomalhaut~b is a neutron star instead of a planet, located in the near fore- or background of the Fomalhaut disk, is strongly constrained by our non-detection. Possible neutron star models fitting the observational data have to be extremely cool ($\lesssim 90\,000$~K) and nearby ($\lesssim 13$~pc). An alternative explanation that the emission seen from Fomalhaut~b is reflected light from the central star is possible, and fits the optical detection in the blue better than the neutron star models which over-estimate the blue emission compared to the observations. Further observations in the UV and EUV would likely be able to clarify the mysterious nature of Fomalhaut~b.

\section{Acknowledgements}
The scientific results reported in this article are based on observations made by the Chandra X-ray Observatory. Support for this work was provided by the National Aeronautics and Space Administration through Chandra Award Number GO6-17056X issued by the Chandra X-ray Observatory Center, which is operated by the Smithsonian Astrophysical Observatory for and on behalf of the National Aeronautics Space Administration under contract NAS8-03060.

\end{document}